\documentclass[twocolumn,amsmath,amssymb,nofootinbib,floatfix,prl]{revtex4}

\usepackage{graphicx}
\usepackage{bm, bbm}      

\begin{document}

\title{``Wormhole'' geometry for entrapping topologically-protected
qubits in non-Abelian quantum Hall states and probing them with
voltage and noise measurements}

\author{
Chang-Yu Hou and Claudio Chamon
}
\affiliation{
Physics Department, Boston University,
590 Commonwealth Ave., Boston, MA 02215, USA
}

\date{\today}

\begin{abstract}

We study a tunneling geometry defined by a single point-contact
constriction that brings to close vicinity two points sitting at the
same edge of a quantum Hall liquid, shortening the trip between the
otherwise spatially separated points along the normal chiral edge
path. This ``wormhole''-like geometry allows for entrapping bulk
quasiparticles between the edge path and the tunnel junction, possibly
realizing a topologically protected qubit if the quasiparticles have
non-Abelian statistics. We show how either noise or simpler voltage
measurements along the edge can probe the non-Abelian nature of the
trapped quasiparticles.

\end{abstract}
\maketitle

Perhaps the largest challenge for constructing a functioning quantum
computer is how to control the loss of coherence in quantum mechanical
systems. There are several different fronts being pursued for
attacking this problem, one of these being the idea of topological
quantum computation~\cite{Kitaev,Freedman}. Decoherence
originates from the contact of the system with its external
environment, and the resilience of topological quantum computation to
decoherence is rooted at degeneracies that cannot be lifted by any
local perturbation or disturbance by the environment. These
degeneracies are associated with topology and not to symmetries, and
were originally proposed by Wen~\cite{Wen} in the context of the
fractional quantum Hall (FQH) effect.

An example of a system with topological degeneracies in its excited
states is the fractional quantum Hall state at $\nu=5/2$, which is
believed to contain quasiparticle excitations with non-Abelian
statistics, and be described by Pfaffian wavefunctions (paired
states)~\cite{Moore-Read,Greiter-etal} as constructed by Read and
Moore. Numerical calculations (exact diagonalizations) lend strong
support that the Pfaffian states do describe the $\nu=5/2$ FQH
state.~\cite{Morf,Rezayi}. The excited state with $2N$ far-apart
quasiholes is $2^{N-1}$-fold degenerate~\cite{Nayak-Wilczek}, and the
degeneracy should be accurate to exponential order in the
characteristic separation between pairs of
quasiparticles. Hence, by keeping the quasiparticles sufficiently
apart, the environment is ineffective in damaging a given quantum
superposition in the $2^{N-1}$-dim Hilbert space. Computation with
non-Abelian anyons is achieved by braiding anyons around each other,
which leads to unitary operations in the degenerate $2^{N-1}$-dim
Hilbert space~\cite{Freedman-Kitaev}. In other words, unitary
operators in the degenerate Hilbert space are associated to the
elements of the braid group acting on the $2N$ particles.


\begin{figure}
\vspace{.5cm}
\includegraphics[angle=0,scale=0.6]{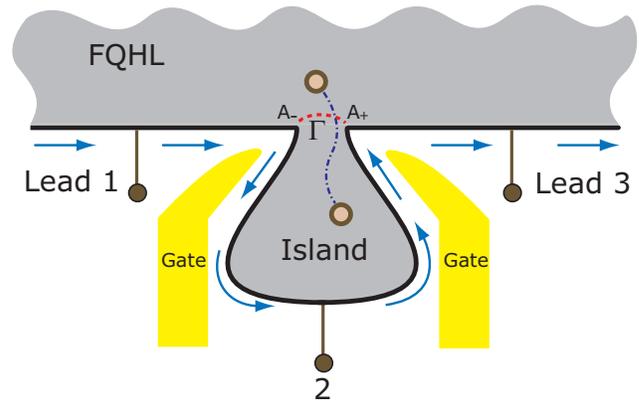}
\caption{A single point-contact pinches a small island out of a FQH
liquid, allowing for the entrapment of quasiparticles inside. The
tunnel junction allows quasiparticles to go from $A_-$ to $A_+$
through the ``wormhole'' and travel faster than the edge velocity that
would take them along the edge path. If the number of quasiparticles
is odd and their statistics non-Abelian, a qubit is stored non-locally
between a quasiparticle inside and a partner outside the
island. Measurements of the differences between voltages $V_1,V_2$ and
$V_3$ with probes at leads $1,2,3$ contain information on the
quasiparticle statistics, which are manifest both in the voltage drops
and their noise correlations.
\label{fig:wormhole}
}
\end{figure}

Certainly, the first step in order to actually realize such quantum
computing scheme is to find a candidate physical system that realizes
non-Abelian statistics. Recently, there have been a few concrete
proposals for experimental probes to investigate the possible
manifestations of non-Abelian statistics in the $\nu=5/2$ FQH
state. One of them explores noise correlations~\cite{Bena}, and four
of them~\cite{Fradkin98,Sarma-etal,Stern-Halperin,Bonderson-etal1} are
interferometric in nature, using a geometry originally proposed for
the detection of Abelian statistics~\cite{CFKSW}. There have been also
proposals to investigate the non-Abelian nature of the $\nu=12/5$
states with interferometry~\cite{Bonderson-etal2,Chung-Stone}.

In this paper we propose a simple geometry for trapping quasiparticles
and probing non-Abelian statistics through voltage measurements in
quantum Hall systems. The assembly requires manipulating a single
chiral edge of the Hall bar with a single point-contact, a
simplification with respect to the geometries of
Refs.~\cite{Fradkin98,Sarma-etal,Stern-Halperin,Bonderson-etal1,CFKSW,Bonderson-etal2,Chung-Stone}. Non-Abelian
statistics can be probed either by measuring voltage drops at
different locations along the edge of the Hall bar, or by measuring
noise correlations between the voltages at these different
points. While the former is certainly an easier measurement than the
latter, noise measurements of similar complexity have been carried out
almost a decade ago and successfully provided evidence for fractional
charge in quantum Hall systems~\cite{Glattli-etal,Reznikov-etal}.

The geometry we propose is depicted in Fig.~\ref{fig:wormhole}, and
requires a single point contact that brings into close vicinity two
points labeled by $A_-$ and $A_+$ separated by a distance $2a$
measured along the chiral edge path. The tunneling path shortens the
travel between points $A_-$ and $A_+$ as compared to the length
measured along the edge, acting as a ``wormhole'' passage that allows
edge quasiparticles to jump over a length $2a$ along the edge. The
edge segment of length $2a$ between $A_-$ and $A_+$, together with the
tunneling path between these two points, encircles a droplet of bulk
FQH liquid. Within this region, quasiparticles can be trapped. If the
statistics of the quasiparticles are non-Abelian, and their number
inside the bounded region is odd, ``half'' of a topologically protected
qubit is entrapped in the island. The qubit corresponds, if formulated
as in Refs.~\cite{Ivanov,Stern-etal}, to the complex fermion
(occupation states $|0\rangle$ and $|1\rangle$) on the string
connecting one quasiparticle inside the region and its long-distance
partner somewhere outside the island.

To detect whether a topologically protected qubit resides inside the
trap, one can measure voltages along the edge. As shown in
Fig.~\ref{fig:wormhole}, one electrode should be placed before the
island (lead 1), one touching the perimeter of the island (lead 2),
and one past the island (lead 3). We focus on the voltage differences
$V_{12}=V_1-V_2$, $V_{13}=V_1-V_3$, and $V_{23}=V_2-V_3$. (One can
simply use leads 1 and 2 instead, but as we discuss below, the third
lead allows for redundancy and checks) The voltage drop $\langle
V_{12} \rangle$, we will show, should be sensitive to the parity of
quasiparticles inside the island, and be a signature of their
non-Abelian statistics. As opposed to the geometries proposed in
Refs.~\cite{Fradkin98,Sarma-etal,Stern-Halperin,Bonderson-etal1,Bonderson-etal2,Chung-Stone},
there is an absolute baseline for the discrimination of non-Abelions:
there should generically be a voltage drop when tunneling is allowed,
unless the statistics is non-Abelian and there is an odd number of
quasiparticles in the island, in which case $\langle V_{12}
\rangle=0$. Notice that $\langle V_{13} \rangle =0$ should always be
the case, which can be seen as a consequence of zero longitudinal
resistance in the FQH effect, or conservation of charge of the chiral
quasiparticles scattered at the junction.  Noise fluctuations of $
V_{12}(t)$, we show, can also tell about the statistics of the
quasiparticles inside.

\begin{figure}
\vspace{.5cm}
\includegraphics[angle=0,scale=0.45]{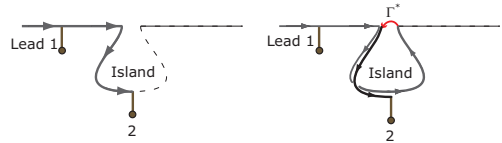}
\caption{Processes up to order ${\cal O}(\Gamma)$ that interfere and
lead to a detectable voltage difference $V_{12}$.
\label{fig:inter}
}
\end{figure}

It is instructive to consider the following argument before going to
the voltage and noise calculation using the edge theory. The leading
processes that interfere to give rise to a voltage differential
between leads 1 and 2 are shown if Fig.~\ref{fig:inter}. For an
incoming many-body scattering state $|\Psi\rangle$,
\begin{eqnarray}
\langle V_{12} \rangle &\propto &
\big|\big|(\openone - i\Gamma^* U^\dagger)|\Psi\rangle\big|\big|^2
-
\big|\big||\Psi\rangle\big|\big|^2
\nonumber\\
&=&
-2\;{\rm Im}\big(\Gamma \langle\Psi|U|\Psi\rangle\big)
\;,
\label{eq:normdiff}
\end{eqnarray}
and the interference term $\langle\Psi|U|\Psi\rangle=0$ when an odd
number of quasiparticles are in the island because the crossing of the
tunneling quasiparticle with the string (a unitary operation $U$)
flips the qubit state with respect to the initial state, so that
$U|\Psi\rangle$ is orthogonal to $|\Psi\rangle$.

The edge theory of $\nu=5/2$ has a charge sector, corresponding to a
$c=1$ free boson theory, and a neutral sector which is one of the
chiral components of the critical theory of the 2D Ising model with
$c=1/2$~\cite{Fradkin98,FNS}. The former part is characterized by the chiral
Luttinger liquid with Lagrangian density,
\begin{equation}
\mathcal{L}_0=-\frac{1}{4\pi} \partial_x \phi (\partial_t+v
\partial_x)\phi\; ,
\end{equation}
where $v$ is the velocity of the (right moving) edge excitations. The
boson fields satisfy the commutation relation, $[\phi(t,x),\phi(t,y)]=
i \pi \; {\rm sgn }(x-y)$. Ignoring the filled Landau levels and
focusing on the half filled one, we will work with $\bar\nu=1/2$
hereafter. Working in units such that $v=1$, the density and current
operators are both given by $\rho(t,x)=\frac{\sqrt{\bar\nu}}{2 \pi}
\partial_x \phi (t,x)$. The Ising sector has three primary fields,
$\openone$(identity), $\sigma$(spin/vortex) of conformal weight
$1/16$, and $\psi$(Majorana fermion) of dimension $1/2$.
The annihilation operator corresponding to the quasiparticle with
fractional charge $e^*=e/4$ can be identified with
$\Psi^{1/4}(t,x)=\sigma(t,x) \;e^{i\sqrt{g}\phi(t,x)}$ for $g=1/8$.

For the geometry shown in Fig.~\ref{fig:wormhole}, the tunneling
operators across the point contact correspond to the annihilation of
a quasiparticle at $x=-a$ and creation at $a$, or vise versa. Thus,
the Lagrangian for the tunneling perturbation is
\begin{equation}
\label{eq:hopping}{L}_t=- \Gamma\; \sigma(t,a)
\sigma(t,-a)\;e^{-i \sqrt{g} \phi(t,a)} e^{i \sqrt{g} \phi(t,-a)}+
{\rm h.c.}
\end{equation}
Hopping of quasiparticles with higher charge is less relevant. The
vertex operators constructed from the charge mode field $\phi$ have
opposite vertex charges, and will correspond to the insertion of
dipoles in the perturbative Coulomb gas expansion.

The quantities that we will focus on are the densities at the three
probe locations at leads $i=1,2,3$: $\rho(t,x_i)$. A few comments are
in order. First, the current and density (which are essentially the
same up to a velocity factor, which we set to unit) are proportional
to the voltage measured at the leads,
$v\left[\rho(t,x_i)-\rho(t,x_j)\right]=\bar\nu
\,\frac{e^2}{h}\;V_{ij}(t)$. For example, a drop or step up in the
voltage $V_{12}$ corresponds to a current that passes through the
``wormhole''. If it were not for the point contact and the particular
overhang geometry, no difference in voltage should occur, according to
a vanishing longitudinal Hall conductance. Second, seen as a many body
scattering problem, the incoming and outgoing chiral states do not
simply differ by a phase, and we thus use the Schwinger-Keldysh
non-equilibrium formalism. Some of the results to first order are not
dependent on using this approach, but other higher order results that
we discuss below do require it.

Within the interaction picture, one defines the scattering operator as
$S(-\infty,-\infty)\equiv\exp[ i \oint ds {L}_t ]$, where the
integral is done over a closed-time path~\cite{Schwinger-Keldysh}. The
expectation value of arbitrary operators, ordered along the Keldysh
contour, are expressed as
\begin{eqnarray}
\label{eq:PerturExpan}\langle T_c\!\left[O_1\cdots O_n\right]\rangle
= \sum_{n=0}^\infty
\langle\Psi|T_c\!\!\left[\frac{( i \oint_c ds {L}_t)^n}{n!}
\,O_1\cdots O_n\right]\!\!|\Psi\rangle
\end{eqnarray}
where $|\Psi\rangle$ is the initial state of the system.

Instead of calculating the density expectation value directly, we can
insert a test charge and compute the expectation of the vertex
operator $\langle e^{i \lambda \phi(t,x)}\rangle$ (constructed using
only the charge mode $\phi$) in perturbation to the desired order and
use the following identity
\begin{equation}
\label{eq:Identity-1}
\langle\rho(t,x)\rangle
=-i \frac{\sqrt{\bar\nu}}{2 \pi} \partial_x
\frac{d}{d\lambda} \langle e^{i \lambda
\phi(t,x)}\rangle\bigg|_{\lambda=0}\; .
\end{equation}
Each order of the perturbation expansion of the test field will
contain the integral over correlation function and has the general
form up to some factors as
\begin{eqnarray}
\oint_c
\prod_{i=1}^{n}\! ds_i
\prod_{j=1}^{m}\! dt_i
\,\langle T_c\!\!
\left
[e^{i \lambda\phi(t,x)}
\prod_{i=1}^n i\mathcal{O}_t(s_i)
\prod_{j=1}^m i\mathcal{O}^\dagger_t(t_j)
\right]\!\rangle ,
\end{eqnarray}
where $n$ and $m$ indicate the perturbation order of the $\Gamma$ and
$\Gamma^*$ respectively and the tunneling operators
$\mathcal{O}_t(s_i)=-\Gamma e^{-i\sqrt{g}\phi(s_i,a)}
e^{i\sqrt{g}\phi(s_i,-a)}$. These matrix elements can be computed
using the Coulomg gas correlation functions for the charge modes
$\phi$~\cite{CFW},
\begin{equation}
\label{eq:Trick-1}
\langle T_c \prod_j e^{i q_j \phi(t_j,x_j)}\rangle
= e^{-\sum_{i>j} q_i q_j \langle T_c \phi(t_i,x_i)
\phi(t_j,x_j)\rangle },
\end{equation}
and the more involved multi-point contour-ordered correlations of the
twist fields $\sigma$.

The lowest non-vanishing contribution to the density/current appears
at the first order:
\begin{equation}
\label{eq:rho}
\langle \rho(t,x)\rangle = \frac{\gamma^*-\gamma}{2 i}\;
\sqrt{g \bar\nu} \,\left[{\rm sgn}(x+a)- {\rm sgn}(x-a)\right],
\end{equation}
where $\gamma$ and $\gamma^*$ include the self-interaction of the
$\phi$-vertex dipole insertion, and the correlation of the twist
fields $\sigma$ that encodes the information on the non-Abelian
statistics of the tunneling quasiparticles:
\begin{equation}
\gamma\equiv \Gamma\;
\langle\Psi|\sigma(s,a)\,\sigma(s,-a)|\Psi\rangle \;\, e^{g\, \langle
T_c( \phi(s,a) \phi(s,-a) )\rangle}.
\end{equation}
The equal-time correlation function of boson field gives a constant
that can be simply absorbed into $\Gamma$. The matrix element
$M=\langle\Psi|\sigma(s,a)\,\sigma(s,-a)|\Psi\rangle$ of the two twist
fields at points $A_-$ and $A_+$ with respect to the quantum state
$|\Psi\rangle$ does capture information on the {\it bulk} FQH state
encircled by the edge perimeter path and the tunneling bridged path
between points $A_-$ and $A_+$. The matrix element can be calculated
using the operator product expansion fusion rules of the $\sigma$
fields, and
has been carried out in this way in Ref.~\cite{Bonderson-etal1} for
the $\nu=5/2$ state (and extended in similar ways to the $\nu=12/5$
state in Refs.~\cite{Bonderson-etal2,Chung-Stone}). $M$ vanishes
if the number of quasiparticles inside the island is odd, while it is
non-zero if the number is even.

It follows from Eq.~(\ref{eq:rho}) using $x_1<-a$ and $x_3>a$ that
$\langle\rho(t,x_1)\rangle=\langle\rho(t,x_3)\rangle=0$ (we just
compute the fluctuations, and subtract a constant term, the zeroth
order term, which in any case cancels out when we calculate density --
and voltage -- differences). Moreover, this result can be shown to
hold order by order in $\Gamma,\Gamma^*$, and it is is simple to
understand why physically. The incoming chiral edge current, at lead
1, cannot be affected by tunneling because of causality, and the
outgoing current, at lead 3, must equal the incoming one by charge
conservation. Notice that this result imply that $\langle
V_{13}\rangle=0$, which is consistent with a vanishing longitudinal
Hall resistance.

However, if one probe is located along the island perimeter at
$-a<x_2<a$, one measures
\begin{equation}
\label{eq:drop-step}
\langle V_{12}(t)\rangle \propto
\langle\rho(t,x_2)-\rho(t,x_1)\rangle=-2\;{\rm Im}(\gamma)\;\sqrt{g
\bar\nu}
\end{equation}
because of tunneling currents through the junction that are
responsible for local deviation from the vanishing longitudinal Hall
resistance. This is the simplest measurement that can probe whether
the bulk quasiparticles trapped within the inside of the island have
non-Abelian statistics. Basically, there should be a detectable
voltage difference $\langle V_{12}\rangle$ whenever the number of
quasiparticles trapped inside the island is even, while there should
be none if the number is odd [${\rm Im}(\gamma)=0$ as $M=0$].

Let us now turn to the behavior of noise correlations, and define
\begin{equation}
\label{eq:Noise}
S_{ij}(\omega)=S_{ji}(-\omega)=
\int_{-\infty}^\infty e^{i \omega t}
\langle\{ \rho(t,x_i), \rho(0,x_j)\}\rangle\; ,
\end{equation}
which can be computed using contour ordered perturbation theory once
we express the expectation value of the anti-commutator as $\langle\{
\rho(t,x_i), \rho(0,x_j)\}\rangle=\sum_\mu \langle
T_c\;\rho_\mu(t,x_i)\, \rho_{-\mu}(0,x_j)\rangle$, where $\mu=\pm$
indicate insertions at the top or bottom branch of the contour,
respectively. From the $S_{ij}(\omega)$, we can obtain the noise
auto-correlations of the quantities $\delta\rho_{ij}(t)\equiv
\rho(t,x_i)-\rho(t,x_j)$ via
\begin{equation}
\label{eq:CrossNoise}
{\mathcal S}_{ij}(\omega)
=S_{ii}(\omega)+S_{jj}(\omega)-S_{ij}(\omega)-S_{ji}(\omega)\;
,
\end{equation}
which are the quantities that are detected if the noise
auto-correlation of the voltage differences $V_{ij}(t)$ are measured.

\begin{widetext}
Carrying out this program, we find to zeroth and first order in the
tunneling amplitude the following:
\begin{eqnarray}
&&{\mathcal S}_{12}^{(0)}(\omega)=2\frac{\bar\nu}{\pi}\;|\omega|
\;\sin^2\left[\frac{\omega(x_2-x_1)}{2}\right]
\\
&&{\mathcal S}_{12}^{(1)}(\omega)= 8 g \bar\nu\; \frac{\gamma+\gamma^*}{2}
\left\{\sin\left[|\omega|(x_2-x_1)\right]\;\sin^2 (\omega a)
+ \sin^2\left[\frac{\omega (x_2-x_1)}{2}\right]\; \sin (2|\omega| a)\right\}.
\end{eqnarray}
\end{widetext}
A simplification occurs if the position of probe 1 is dithered, or
else the edge path length is modulated, or the distance $|x_1-x_2|$ is
large enough for dephasing to occur (here we actually profit from loss
of coherence). In this case, as long as the observation frequencies
are large compared to the inverse of the time of flight through the
length $|x_1-x_2|$, one can average over $x_1-x_2$ and obtain
\begin{eqnarray}
&&\overline{{\mathcal
S}_{12}^{(0)}}(\omega)=\frac{\bar\nu}{\pi}\;|\omega|
\\
&&\overline{{\mathcal S}_{12}^{(1)}}(\omega)= 4 g \bar\nu \;{\rm Re}(\gamma)
\;\sin (2|\omega| a).
\end{eqnarray}
It is noteworthy that $\overline{{\mathcal
S}_{12}^{(1)}}(\omega)={{S}_{22}^{(1)}}(\omega)$, {\it i.e.},
the excess noise comes all from the auto-correlation measured at lead
2.

Notice that the excess noise $\overline{{\mathcal
S}_{12}^{(1)}}(\omega)$ in the voltage difference $V_{12}$ is
proportional to ${\rm Re}(\gamma)$, and hence depends on whether the
number of quasiparticles within the island is even or odd. Moreover,
in this geometry, the voltage difference measurement and the noise are
{\it not} proportional to each other; instead they are {\it in
quadrature}, one proportional to ${\rm Re}(\gamma)$ and the other to
${\rm Im}(\gamma)$. If both are measured, (un)fortuitous cancellations
of one due to destructive interference will be accompanied by a
maximum in the other. Hence, no signal can be detected for both
measurements only if $\gamma=0$. This scheme provides a redundant
check and a strong constraint on the even/odd detection recipe.

Let us turn the discussion to what should happen in the regime of
strong coupling. When tunneling is strong, the incoming edge current
should pass mostly through the ``wormhole'', leaving an isolated
island aside. Within this isolated FQH puddle an integer number of
electrons must reside, and thus only an even number of quasiparticles
is permitted (actually, a multiple of 4, since $e^*=e/4$). Therefore,
if a topological qubit was realized for an odd number of
quasiparticles in the island at weak coupling, it should be
``screened'' at strong coupling. We speculate that this screening
takes place as in the Kondo effect at strong coupling, where the
magnetic impurity is screened by the conduction electrons. The
two-state system corresponding to the qubit can be represented as a
complex fermion shared by a quasiparticle in the island and another
outside, and as the tunneling amplitude gets larger, the string
connecting this pair gets crossed more and and more frequently, and
the qubit is rapidly flipped, and should ``hybridize'' with the edge
quasiparticles and be screened. The connection between the trapped
topological qubit and the Kondo effect seems appealing, but in order
to substantiate this scenario, one must learn how to deal with the
non-Abelian phase factors due to the products of a number of twist
operators $\sigma$, which is certainly a non-obvious problem, and one
that deserves further investigation.




The authors would like to thank Chetan Nayak and Kirill Shtengel for
stimulating and useful discussions. This work is supported by the NSF
grant DMR-0305482.


\end{document}